\documentclass[floatfix,onecolumn,pra,aps]{revtex4-1}

\usepackage{hyperref}
\usepackage{graphicx}  
\usepackage{dcolumn}   
\usepackage{bm}        
\usepackage{amssymb}   
\usepackage{epstopdf}
\usepackage{amsmath}
\usepackage{xspace}
\usepackage{verbatim}
\usepackage{xcolor}
\usepackage{amsmath}

\newcommand{\bra}[1]{\left\langle#1 \right|}
\newcommand{\ket}[1]{\left|#1 \right\rangle}
\newcommand{\up}{|\uparrow\rangle}
\newcommand{\updag}{\langle\uparrow|}
\newcommand{\down}{|\downarrow\rangle}
\newcommand{\downdag}{\langle\downarrow|}
\newcommand{\inner}[2]{\left\langle #1 | #2 \right\rangle}

\begin{document}

\title{Phonon Maser Stimulated by Spin Post-selection}

\author{Vitalie Eremeev}
\email{vitalie.eremeev@udp.cl} 
\affiliation{Facultad de Ingenier\'ia y Ciencias, Universidad Diego Portales, Av. Ejercito 441, Santiago, Chile}
\affiliation{Institute of Applied Physics, Academiei str. 5, MD-2028 Chi\c{s}in\v{a}u, Moldova.}

\author{Miguel Orszag}
\email{miguel.orszag@umayor.cl}
\affiliation{Centro de Optica e Informaci\'on Cuantica, Universidad Mayor, camino la Piramide 5750, Huechuraba, Santiago, Chile}
\affiliation{Instituto de F\'isica, Pontificia Universidad Cat\'olica de Chile, Casilla 306, Santiago, Chile.}

\date{June 15, 2019}

\begin{abstract}
In a  sequence of single spins interacting longitudinally with a mechanical oscillator, and using the micromaser model with random injection, we show that after an appropriate post-selection of each spin, a phonon laser analog with Poisson statistics is created with nearly perfect coherence, evidenced by the second order coherence function that goes asymptotically to one.
The non-linear gain of the system depends crucially on the properly post-selected spin state as well as the pump. 
Our model and results suggest that the mechanism of interaction followed by a post-selected state or partial trace (common in laser/maser theory) of the spins may create the coherent vibrational radiation. However, for situations where the mechanical losses are high and it is imposible to decrease these, then the heralded post-selection can be the only resource to get phonon lasing if compared to the partial trace operation. These ideas and results may be useful for further theoretical and technical developments.

\end{abstract}

\maketitle

\section{Introduction}

A quantum system coupled to the motion of a micro- or nano-fabricated mechanical oscillator has become of pivotal importance in realizing quantum technological tasks. Myriad applications such as cooling of phonon modes to enhance sensing \cite {Al,Cl,Ru}, exploration of quantum effects \cite {Chan,Tuf, Ceb, Cohen, Rabl} and others have been realized. Over the last decades, besides cooling or creating quantum vibrational states \cite {Jin, Rao, Monte1, Monte2}, there has been a great interest in the research of generating a phonon laser effect.
Some theoretical proposals include vibrational amplification of a single trapped ion \cite {Wall}, nano-mechanical analog of a laser \cite {Ba}, phonon laser effects in nanomagnets \cite {Ch}.
On the experimental side, there have been several results related to ultrasonic pulses by maser action \cite {Tuc}, stimulated emission of phonons in ruby \cite {Bron, Hu}, stimulated emission in an acoustic cavity \cite {F}.
More recently, a phonon laser was realized using a single trapped ion and two laser beams \cite {Vahala}, in a microcavity system coupled to a radio frequency mechanical mode \cite {G}, in an electromechanical resonator \cite {Ma}, and in an optomechanical system with a silica nanosphere levitating in an optical tweezer \cite {Pet}.

From the history of the L(M)aser development we know that many kind of systems and concepts were proposed to reach the ultimate effect - Amplification of Radiation by Stimulated Emission. Hence, this impressive experience drove the community to find the feasible mechanisms for a phonon laser analog, which presents nowadays new technological challenges. In the spirit of this idea, we propose here to take into account the richness of the physical effects discovered in the last years by using the hybrid systems, usually composed by the quantum systems and mesoscopic object as mechanical oscillator/resonator, metamaterial, etc. \cite {Kur, Asp, Alba} Additionally to the richness and effectiveness of the hybrid system, we propose to include also the advantages of the post-selective measurement, proved to play a key role as found in some recent theoretical and experimental investigations  \cite {Rao, Coto, Monte1, Monte2, Car, Hal, Chrz, Walk}.

Measurements in quantum mechanics are usually described by the interaction of a system we want to measure and the measurement apparatus in such a way that the modification of the probe state depends on the value of the observable.
If this interaction is strong, and the apparatus is represented by a narrow wave-function, as compared to the spectrum gaps of our observable, we get the usual von Neumann scenario, where the state of the system is strongly modified by the measurement.
On the other hand, Aharonov \textit{et.al.} \cite {Ah} proposed first the idea of weak measurements combined with the pre- and post-selections, where the measurement apparatus was represented by a state with a very large uncertainty, when compared with the typical distance between the eigenvalues of a given observable.
As a result, one can get, under certain conditions, an amplification of a small effect, and the final state of our system is hardly modified at all.
Although there were some claims that the amplification obtained from weak interaction followed by post-selection had a classical nature \cite {Fer}, it was later proven of quantum origin \cite {Mun}.
There is a large amount of literature in connection to successfully realized post-selection measurements in different setups \cite {Feiz, Les, Viza} and weak value amplification (WVA), on theoretical and experimental grounds. In particular, WVA has been used to estimate small parameters like precision frequency measurements with interferometric weak values \cite {St}, enlarge birefringent effects \cite {Ric}, or sensitive estimation of angular rotations of a classical beam, getting an amplification as big as one hundred \cite {Mag}, just to mention a few.

Finally, WVA has become crucial to observe directly the wave function and trajectory in a two-slit experiment \cite {Koc}, and single photon amplification, both as a non-linear effect or in an optomechanical interaction \cite {Hal, Car}. These ideas will prove useful for the proposed phonon laser model in a spin-mechanical system pumped by a combination of randomly injected spins and post-selection of particular spin states. As demonstrative examples we consider the cases where the lasing is realized within the mechanisms of post-selection with successful and unsuccessful readouts, also with the partial tracing operation. On the other hand, the most important result we want to emphasize in this work is that by using the advantage of the controled post-selection protocols \cite {Rao, Walk, Chrz, Wang} it is possible to amplify considerably the effect of lasing and so, for situations where the mechanical losses are high enough, then the heralded post-selection with the optimal initial and final states (nearly orthogonal) is the only resource to realize phonon lasing if compared to the partial trace operation under the optimal initial state, which is spin ''up''  or ''down'' in this case.
To the best of our knowledge, such a phonon lasing model is an original one, particularly under the framework of the micromaser theory together with the post-selective measurement of the spin interacting non-resonantly with a mechanical oscillator. We point out that this proposal comes as a logical continuation of our previous work \cite {Monte1, Monte2}, where we demonstrated several interesting effects as a consequence of the pre- and post-selective measurements applied for a hybrid system with the spin-mechanical longitudinal coupling \cite {Jin, Rao}, i.e. interaction without energy exchange.

\section{METHODS}
\subsection{Model of phonon maser assisted by post-selection}
In this work we present as a theoretical model, an alternative scheme for a phonon laser, and will show that in some cases the pre- and post-selective measurements play an important role to obtain and amplify the lasing effect. To understand better the impact of the post-selection on the lasing effect we will present the comparison with the case of \textit{spin tracing} operation, which is the standard method in the laser theory. For our phonon lasing proposal we are inspired on the one-atom maser (micromaser) like model \cite {Meschede, Filip, Rempe} by applying it to a hybrid system like \cite {Monte1}, where a mechanical oscillator interacts longitudinally with a spin during a fixed short time, $\tau$. At the end of the interaction, we post-select a state of the spin, with a certain probability. If this process is successful, the oscillator relaxes a longer time, i.e. $\Delta t \gg \tau$, under the action of the thermal bath until the next spin is ready to interact. These processes of interaction, post-selection and relaxation continue until the mechanics reaches a phonon steady-state with a nearly Poisson distribution plus phonon amplification, hence showing phonon lasing.
\begin{figure}[t]
\centering
	\includegraphics [scale = 0.7] {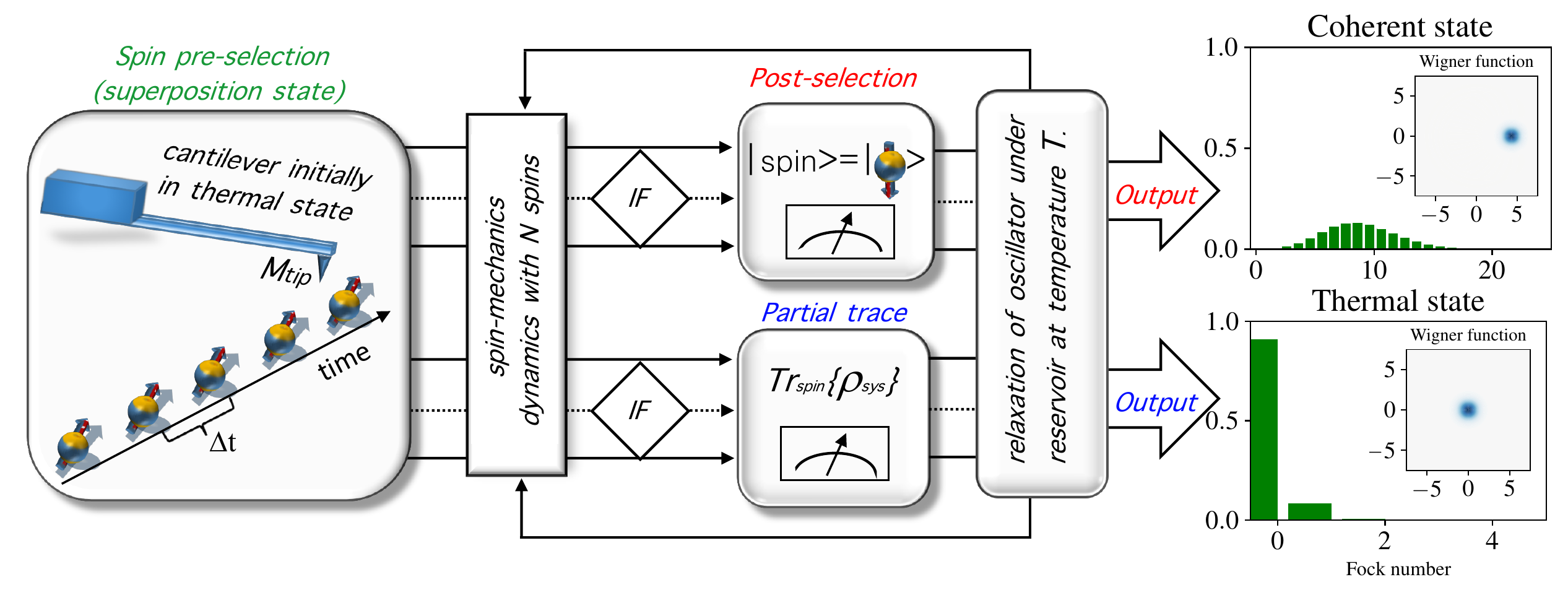}
	\caption{Artistic sketch of the \textit{phonon maser} model, where one spin is approaching to the magnetic tip ($M_{tip}$) at the time interval $\Delta t$ and interacts longitudinally with a mechanical oscillator during the time $\tau$. The damping mechanisms to thermal environments are present for the oscillator at the rate $\kappa$ in units of the oscillator's frequency, $\omega_m$. By this scheme we want to evidence the two possible outputs which depend crucially on the type of spin measurement, used at an intermediate step of the micromaser model.}
	\label{fig:model}
\end{figure}

The main idea is pictorially represented in Fig.(\ref{fig:model}) with a sequence of spin qubits coupled longitudinally \cite {Jin, Rao, Monte1} to a mechanical oscillator, with specified pre- and post-selected spin states.
 This elementary system is described in the interaction picture by (with $\hbar = 1$)
 \begin{equation}
  \hat{H} = \hat{b}^\dag \hat{b} - \lambda \hat{\sigma}_z (\hat{b}^\dag  +  \hat{b}), \label{eq:Hint}
\end{equation}
where $\lambda = \lambda_0/\omega_m$ is the scaled coupling strength, $\lambda_0$ the direct spin-mechanical coupling interaction and $\omega_m$ the oscillator frequency; $\hat{b}$ stands for the annihilation bosonic operator.
We assume the oscillator to be initially in a thermal state $\hat{\rho}_m(0)\equiv\frac{1}{2\pi\bar{n}_0} \int \text{d}^2\beta e^{-\frac{|\beta|^2}{\bar{n}_0}} \left| \beta \right\rangle \left \langle \beta \right|$, with $\beta=r\exp[i\phi]$ representing the amplitude of the coherent state; We additionally assume that we have a low initial phonon number, i.e. $\bar{n}_0 < 1$. The spin is pre-selected in the state $\hat{\rho}_s(0)=(\up + \down)(\updag + \downdag)/2 $, and the initial state of the entire system (oscillator + spin) reads as $ \hat{\rho}(0)=\hat{\rho}_m(0)  \otimes \hat{\rho}_s(0) $.

 The main task of our lasing protocol is to evolve the pre-selected spin under the interaction energy for a given time $\tau$, and to post-select the spin in a target state  $\ket{\psi_\text{t}} = \cos\theta \up + \sin\theta \down$. The dynamics of the spin-mechanics is calculated by using the unitary time evolution operator $\hat{U}(\tau)=\exp[ \lambda \hat{\sigma}_z (\eta \hat{b}^\dag  -  \eta^* \hat{b})] \exp[ -i\hat{b}^\dag\hat{b} \tau ]$, derived in \cite {Monte1} from $\hat{H} $ with $\eta=1-e^{-i\tau}$; for more details about the calculation of the unitary operator for a similar Hamiltonian see \cite {Bose}. Therefore, the evolved mechanical state after the spin  post-selection, reads
 \begin{equation} 
 \hat{\rho}_m(\tau)=\bra{\psi_\text{t}} \hat{U}(\tau) \hat{\rho} (0) \hat{U}^\dagger(\tau)  \ket{\psi_\text{t}}.  \label{eq:rho osc}
 \end{equation}
 After a straightforward calculation, one gets the normalized mechanical density operator
 \begin{eqnarray}
 \label{eq:rho_evol}
 \hat{\rho}_m(\tau) &= & \frac{1}{\mathcal{N}} \left(  \cos^2\theta \mathit{D}(\lambda\eta) e^{-i\hat{b}^\dag\hat{b} \tau}  \hat{\rho}_m(0) e^{i\hat{b}^\dag\hat{b} \tau}\mathit{D}(-\lambda\eta) + \sin^2\theta \mathit{D}(-\lambda\eta) e^{-i\hat{b}^\dag\hat{b} \tau}  \hat{\rho}_m(0) e^{i\hat{b}^\dag\hat{b} \tau}\mathit{D}(\lambda\eta) \right. \\  \nonumber
  &+& \left. \frac{\sin 2\theta}{2} \left[ \mathit{D}(\lambda\eta) e^{-i\hat{b}^\dag\hat{b} \tau}  \hat{\rho}_m(0) e^{i\hat{b}^\dag\hat{b} \tau}\mathit{D}(\lambda\eta)+ h.c. \right] \right), 
 \end{eqnarray}
 where $\mathcal{N}= \left( \cos^2\theta + \sin^2\theta + \frac{\sin 2\theta}{2} \text{Tr} \left\{ \hat{\rho}_m(0) e^{i\hat{b}^\dag\hat{b} \tau} \left[ \mathit{D^2}(\lambda\eta) +  \mathit{D^2}(-\lambda\eta ) \right] e^{-i\hat{b}^\dag\hat{b} \tau}  \right\} \right)$ and $\mathit{D}(\alpha)$ is the usual displacement operator. For example, for $\theta=\pi/2$ one has $\mathcal{N}=1$.

\subsection{Maser Master Equation}
 Next, we make use of the well known \textit{Micromaser Model}. From that viewpoint, it is clear that Eq.(\ref{eq:rho_evol}) represents the gain corresponding to a single spin of the micromaser master equation (ME), as described for example in \cite{Orszag} by using the quantum theory of the laser. So the gain part can be written as $ \hat{\rho}_m(\tau)\equiv \hat{M}(\tau) \hat{\rho}_m(0)$, where $\hat{M}$ is a superoperator generally deduced from the system's Hamiltonian, in our case considering Eq.(\ref{eq:Hint}). Next we fix the interacion time and the post-selected state, e.g. $\tau=\pi$ and $\theta=\pi/2$, so  $ \hat{M}(\tau) \hat{\rho}_m(0)=\mathit{D}(-\lambda\eta) e^{-i\hat{b}^\dag\hat{b} \tau}  \hat{\rho}_m(0) e^{i\hat{b}^\dag\hat{b} \tau}\mathit{D}(\lambda\eta)$. 
 The full ME, considering the gain term assisted by spins injected at the rate $r$ (see Appendix A) and the loss term, reads \cite{Orszag}
 \begin{equation} 
 \dot{\hat{\rho}}_m(t)=r \left( \hat{M}(\tau) -1 \right)  \hat{\rho}_m(t) + \mathcal{L}\hat{\rho}_m(t),  \label{eq:ME}
 \end{equation}
  where $\mathcal{L}\hat{\rho}_m(t)=\kappa (1+\bar{n}_0) \left( \hat{b}\hat{\rho}_m\hat{b}^{\dagger}-\frac{1}{2}
 \hat{\rho}_m\hat{b}^{\dagger}\hat{b}-\frac{1}{2}\hat{b}^{\dagger}\hat{b}\hat{\rho}_m\right)+\kappa \bar{n}_0 \left( \hat{b}^{\dagger}\hat{\rho}_m\hat{b}-\frac{1}{2}
 \hat{\rho}_m\hat{b}\hat{b}^{\dagger}-\frac{1}{2}\hat{b}\hat{b}^{\dagger}\hat{\rho}_m\right) $ is the standard Lindbladian describing the field decoherence; here $\kappa$ is the rate of the phonon damping to the bath with $\bar{n}_0=(\exp[\hbar \omega_m/k_BT]-1)^{-1}$ phonons (the mechanics is initially in the same state) at temperature $T$ and $r=1/\Delta t$ with $\Delta t$ being the time between two consecutive spin-oscillator interactions satisfying the necessary condition for the maser model  $\Delta t\gg\tau$. 

 A more general model includes pump statistics (see Appendix A), but here we assume a random arrival and measurement of the spins interacting with the oscillator, that corresponds to $p\to0$ in the micromaser notation. Such an approach for our model is reasonable and practical from experimental point of view. The argumentation is that the preparation (pre-selection) and measurement (post-selection) are inherently probabilistic processes, hence the random arrival (incoherent pump) of the spins will adequately "simulate the physics" of the probabilistic events involved in the model.
 
\section{RESULTS}
\subsection{Scenario for heralded successful post-selections.}
In the following we consider the particular case, when the spin post-selection is successful after each spin-mechanical interaction, i.e. the readout gives the state $\ket{\psi_\text{t}}$ if the previous readout was successful, as result the mechanical density operator defined by Eq.(\ref{eq:rho_evol}) evolves within the maser ME, Eq.(\ref{eq:ME}). Such a scenario looks more like a theoretical idealization, nevertheless it is not only a theoretical idea \cite{Rao, Walk}, but can be implemented in some experimental setups \cite{Rao, Chrz, Wang}, known as heralded control of the post-selective measurement. Hence we consider this case as useful for comparison with more common experimental situations discussed further.
In order to witness the laser effect we calculate the properties such as phonon statistics, second order coherence function, $g^{(2)}(0)$, and the linewidth. 

\textit{Average phonon number and distribution function.} From the ME (\ref{eq:ME}), we can deal with the dynamics and solve a Fokker-Planck equation, using Glauber's $P$-distribution, which we solved analytically (see Appendix B), obtaining for the mean phonon number, (\ref{eq:B3}), the following result
\begin{equation} 
\left\langle \hat{n}(t) \right\rangle = \bar{n}_0+\frac{16 \lambda^2 r^2}{ \kappa^2} \big ( 1- \exp[-\kappa t/2]  \big)^2.  \label{eq:n_t}
\end{equation}
Hence, the steady-state ($t \to \infty$) average phonon number is $\bar{n}_{SS}=\bar{n}_0+16\lambda^2r^2/\kappa^2$.
    \begin{figure}[h]
    \centering
 	\includegraphics[scale = 0.4] {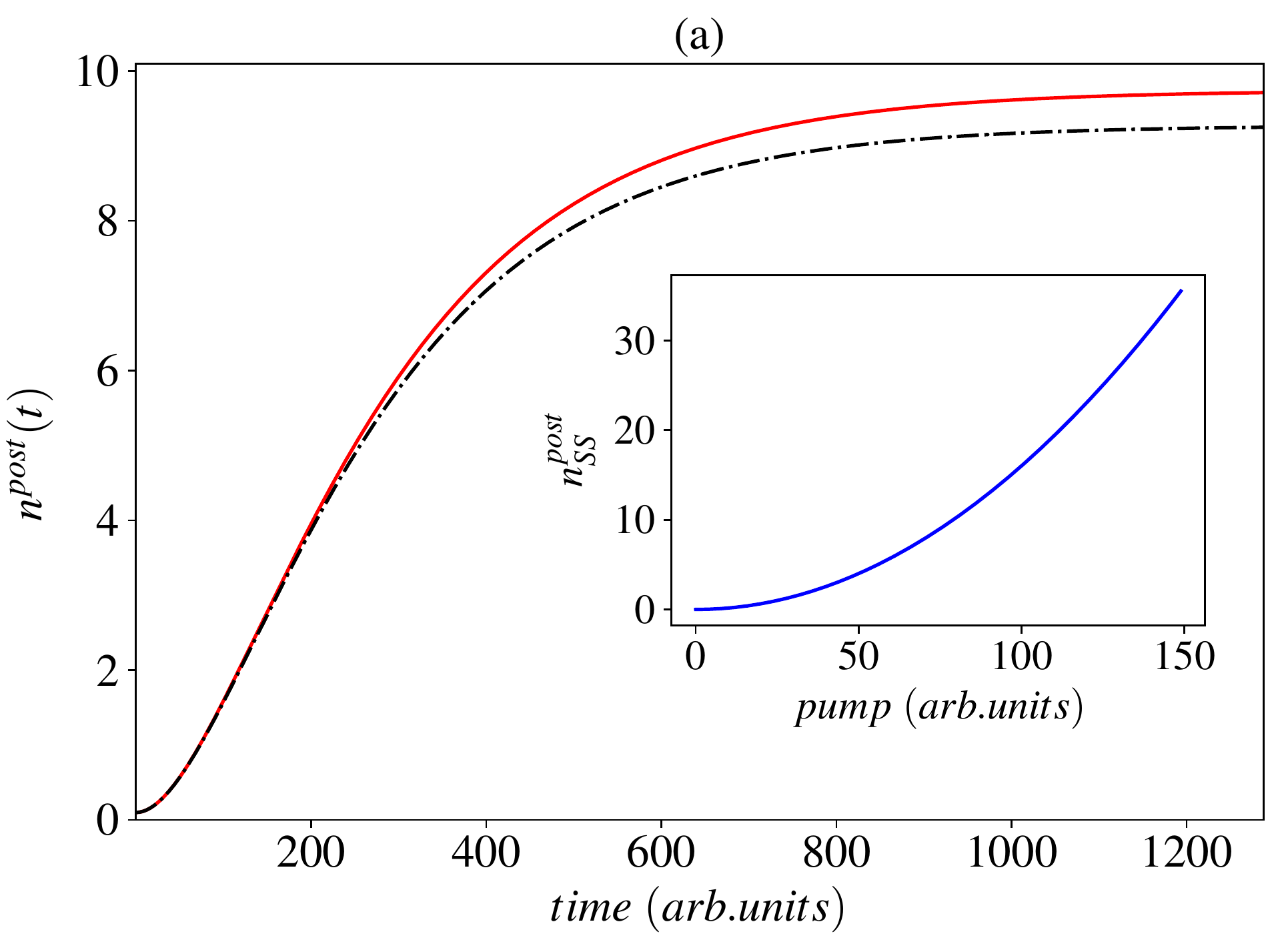}
 	\includegraphics[scale = 0.38] {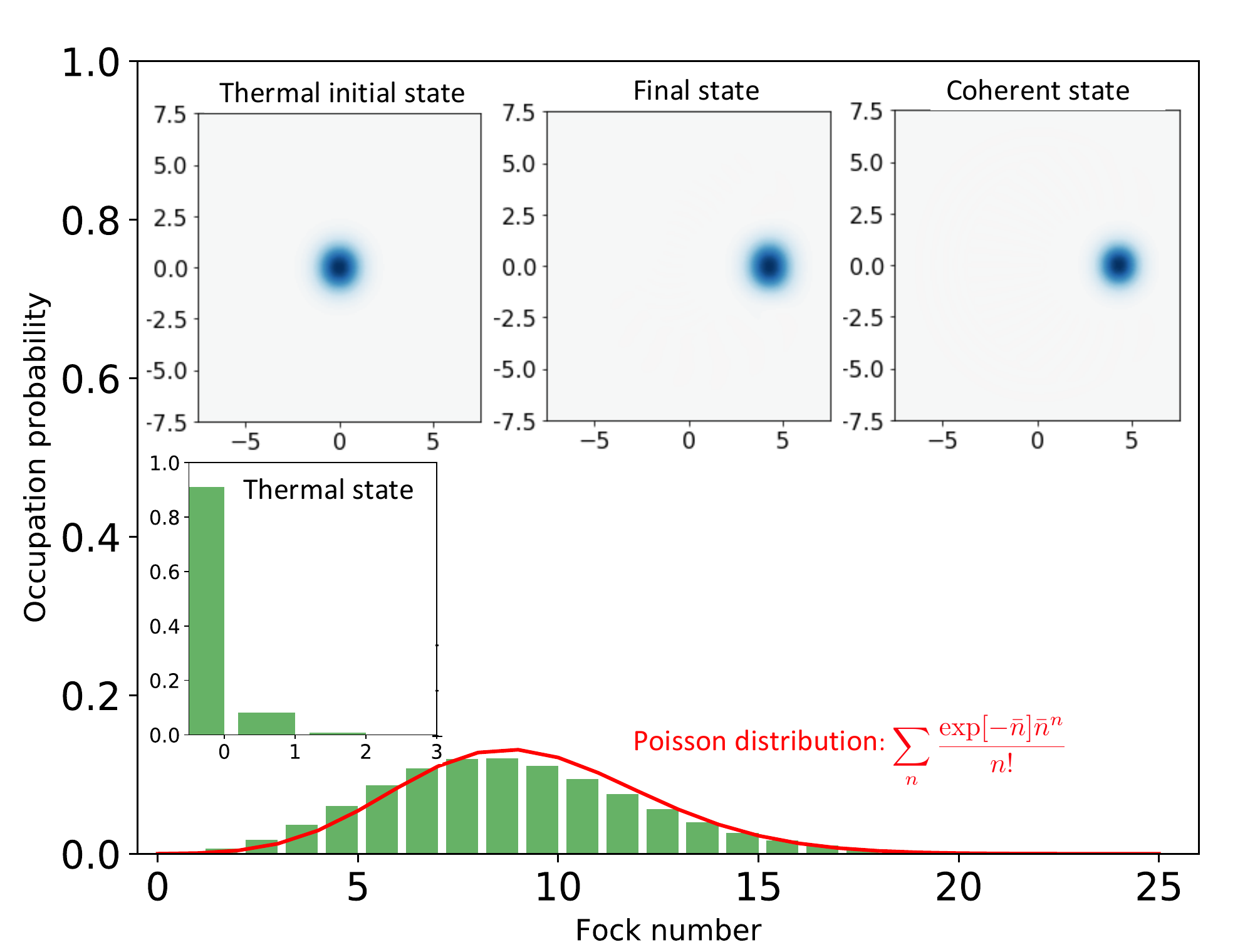}
 	\caption{(Left panel) Time evolution of the average phonon number calculated: i) using analytical expression Eq.(\ref{eq:n_t}) which converge to the steady-state  $\bar{n}_{SS}=9.7$ (red dashed) and ii) using numerical simulation of the ME (\ref{eq:ME}) which converges to the steady-state  $\bar{n}_{SS}\approx9.3$ (dot-dashed). (Inset) Steady state phonon number vs. pump $\propto r^2$, as in Eq.(\ref{eq:n_t}). \\
 		(Right panel) Probability distribution function evidencing the steady-state solution (green bars) and compared to the standard Poisson distribution for the same mean value (red line). (Inset)  Wigner function plots and the thermal initial distribution. The model parameters are: $\bar{n}_0=0.1$, $\lambda=0.001$, $\kappa=0.01*\lambda$ and $\Delta t=41*\tau$, for $\tau=\pi$.}
 	\label{fig:nmed}
 \end{figure}
 
Thereafter, the phonon probability distribution function is calculated for the steady-state solution, i.e. in Eq.(\ref{eq:ME}) considering $\dot{\hat{\rho}}_m=0$, and reads (see Appendix B)
\begin{equation} 
P(n)\equiv \hat{\rho}_{m_{ n,n}} = \frac{1}{\pi \bar{n}_0 n!} \int \text{d}^2 \beta \vert\beta\vert^{2n} e^{-\vert\beta\vert^2 - \frac{\vert\beta-\bar{\beta}\vert^2}{\bar{n}_0}}, \label{eq:prob}
\end{equation}
where $\vert\bar{\beta}\vert^2\equiv \bar{n}_{SS}-\bar{n}_0$. By numerical calculation of Eq.(\ref{eq:n_t}) we plot the time evolution of the average phonon number and the respective phonon distribution function, Fig. (\ref{fig:nmed}).
In the Fig. (\ref{fig:nmed}a), we show the growth of the average phonon number as a function of time, which for longer times, saturates to a steady state value. On the other hand, in Fig. (\ref{fig:nmed}b), we present the steady state phonon statistics obtained and compared to an exact Poisson distribution with the same phonon mean value. We observe that the two distributions are very similar, indicating that the state of the phonons is nearly coherent.

\textit{Second order coherence function.} In order to witness the degree of coherence of the laser emission we evaluate the second-order correlation function, $g^{(2)}$. Therefore, defining a generating function $Q(s)=\sum_{n=0}^{\infty}(1-s)^n P(n)$, with $ P(n)=\hat{\rho}_{m_{ n,n}}$ as in Eq.(\ref{eq:prob}).
It is simple to prove (see e.g. \cite{SZ}) that 
$ g^{(2)}(0)=\frac{1}{\langle n\rangle} \frac{d^{2}Q}{ds^{2}} \mid
_{s=0}=\frac{\langle n(n-1)\rangle }{\langle n\rangle ^{2}}$, where $\langle n\rangle=-\frac{d Q}{ds}$.

Hence, after a simple but rather long calculation, one gets the final expression
\begin{equation}
g^{(2)}(0)=\frac{2\bar{n}_0^{2}+4\beta _{1}^{2}\bar{n}_0+\beta _{1}^{4}}{\bar{n}_0^{2}+2\beta _{1}^{2}\bar{n}_0+\beta _{1}^{4}}. 
	\label{eq:eqg2}
\end{equation}

\vspace{10pt}
\textit{Linewidth.} Another qualitative witness of the laser emission is its linewidth, which in fact results from the intrinsic quantum nature of the lasing. Hence, to complete our analysis of the phonon maser in the following we calculate the analytical expression of the linewith. The Fokker-Planck equation, Eq.(\ref{eq: A1}), can be written in polar coordinates
with $\beta =r\exp i\theta$ \cite{SZ}.
Therefore, for the steady state that corresponds to the regime with $\frac{\partial }{\partial r}=0$ one has: $\frac{\partial P}{\partial t}=0=\frac{\mathcal{D}}{2}\frac{\partial^{2}P}{\partial \theta ^{2}}$, where 
\begin{equation}
\mathcal{D}=\frac{\kappa \bar{n}_{0}}{2 \bar{n}_{SS} }=\frac{\kappa \bar{n}_{0}}{2(\bar{n}_{0}+16\lambda ^{2}r^{2}/\kappa ^{2})}
\label{eq:D1}
\end{equation}

is the linewidth for the  phonon maser steady state regime.

\begin{figure}[h]
 \centering
 	\includegraphics [width=0.5\linewidth] {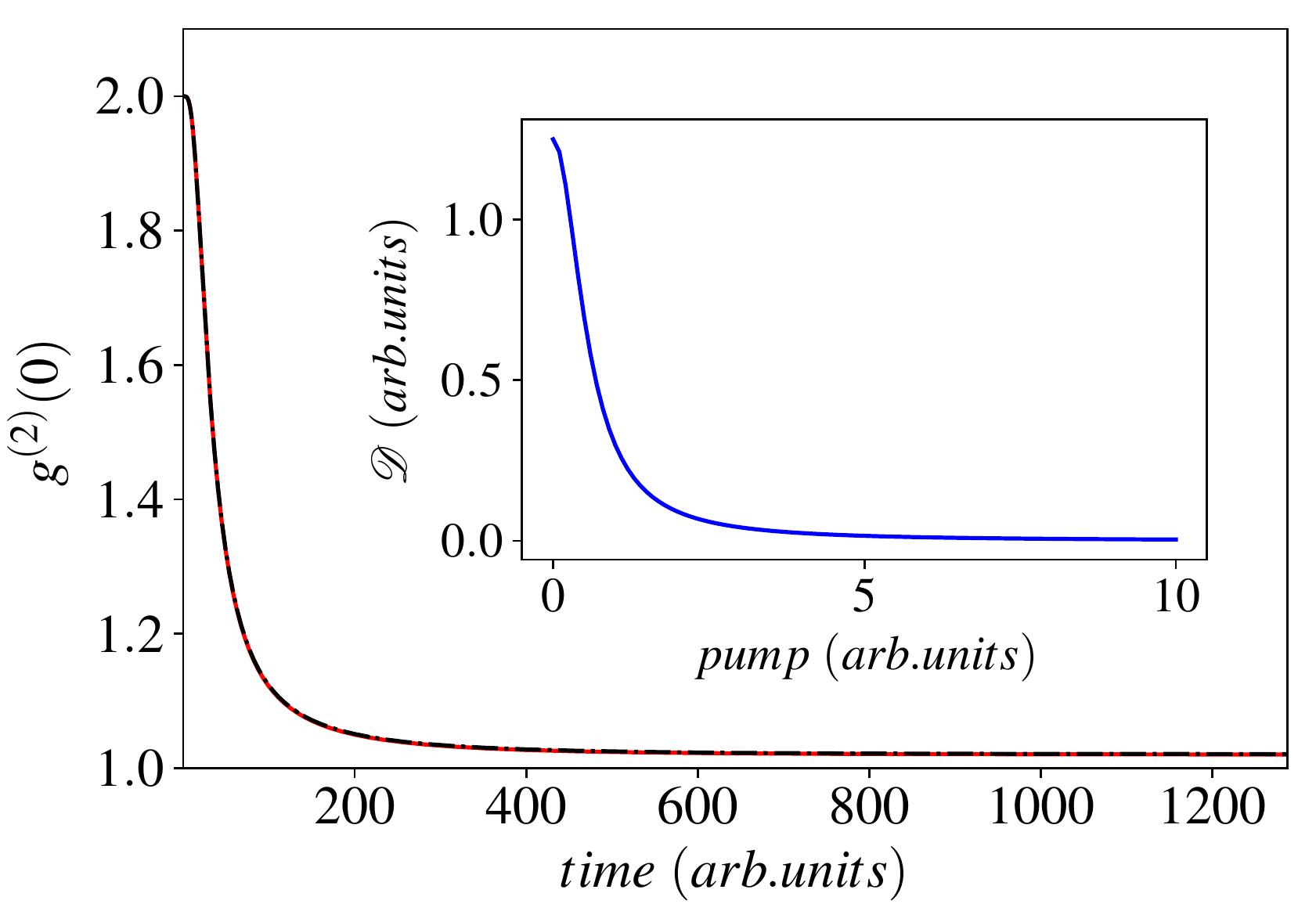} 
 	\caption{Second-order coherence degree, evidencing the formation of coherent state when $g^{(2)}(0)\to 1$. Here we compare: i) analytical expression Eq.(\ref{eq:eqg2}) (red dashed) and ii) numerical solution resulted from the ME (\ref{eq:ME}) (black dot-dashed). Both curves fit perfectly.
 		(Inset) Linewidth vs. pump $\propto r^2$, as in Eq.(\ref{eq:D1}) }
 	\label{fig:g2}
 \end{figure}

As result, in the main plot of Fig. (\ref{fig:g2}) we present the analytical solution Eq.(\ref{eq:eqg2}) for the second order coherence versus time which evidences a very good agreement with the numerical simulation (red dashed) of the ME (\ref{eq:ME}). As shown, the phonon steady state correspond to a coherent state, i.e. $g^{(2)}(0)\to 1$ for long times. In the Inset of the same figure we plot the linewidth as in Eq.(\ref{eq:D1}), which is observed to behave similarly as in the standard laser model \cite{SZ}. 
 
One could think that instead of post-selecting the spins, we could average over the spin variables, that is, to perform a \textit{partial trace} and study the evolution of the system. Thus, we readily get the evolved mechanical density matrix
\begin{equation}
\label{eq:rho_tr}
\hat{\rho}_m(\tau) \equiv Tr_s[\hat{U}(\tau) \hat{\rho} (0) \hat{U}^\dagger(\tau)] =\mathit{D}(\lambda\eta) e^{-i\hat{b}^\dag\hat{b} \tau}  \hat{\rho}_m(0) e^{i\hat{b}^\dag\hat{b} \tau}\mathit{D}(-\lambda\eta) + \mathit{D}(-\lambda\eta) e^{-i\hat{b}^\dag\hat{b} \tau}  \hat{\rho}_m(0) e^{i\hat{b}^\dag\hat{b} \tau}\mathit{D}(\lambda\eta).
\end{equation}
In the micromaser language, the above expression corresponds to the gain part of the ME (\ref{eq:ME}). After performing the same calculation as in the post-selected version, we arrive to the following conclusions:
(A) The final state of the mechanical oscillator differs very little from the initial thermal one (see Inset of Fig. (\ref{fig:nmed}b), and
(B) There is no indication of any lasing effect. As a matter of fact, there is hardly any change in the average phonon number, during its time evolution. Starting from $\bar{n}_0$, the evolution results in an increase of less than $5\%$, behavior that shows no amplification of phonons.

Therefore, the main conclusion in this scenario is that for the conditions considered above there exists lasing only for the mechanism of successful post-selections, i.e. not including the failures in the ME. Such kind of successful post-selection mechanism is applied in the experimental setups with heralded control \cite {Rao, Chrz}, where the post-selection measurements are used iteratively and hence the sequence of successful readouts are important. On the other hand, we found that under the approaches of \textit{spin tracing} and post-selections with failures there is no lasing observed under the conditions of this scenario, i.e. the initial and final spin states.

 \subsection{Scenario for the imperfect post-selection including the failures.}
\begin{figure}[t]
 	\includegraphics[scale = 0.5] {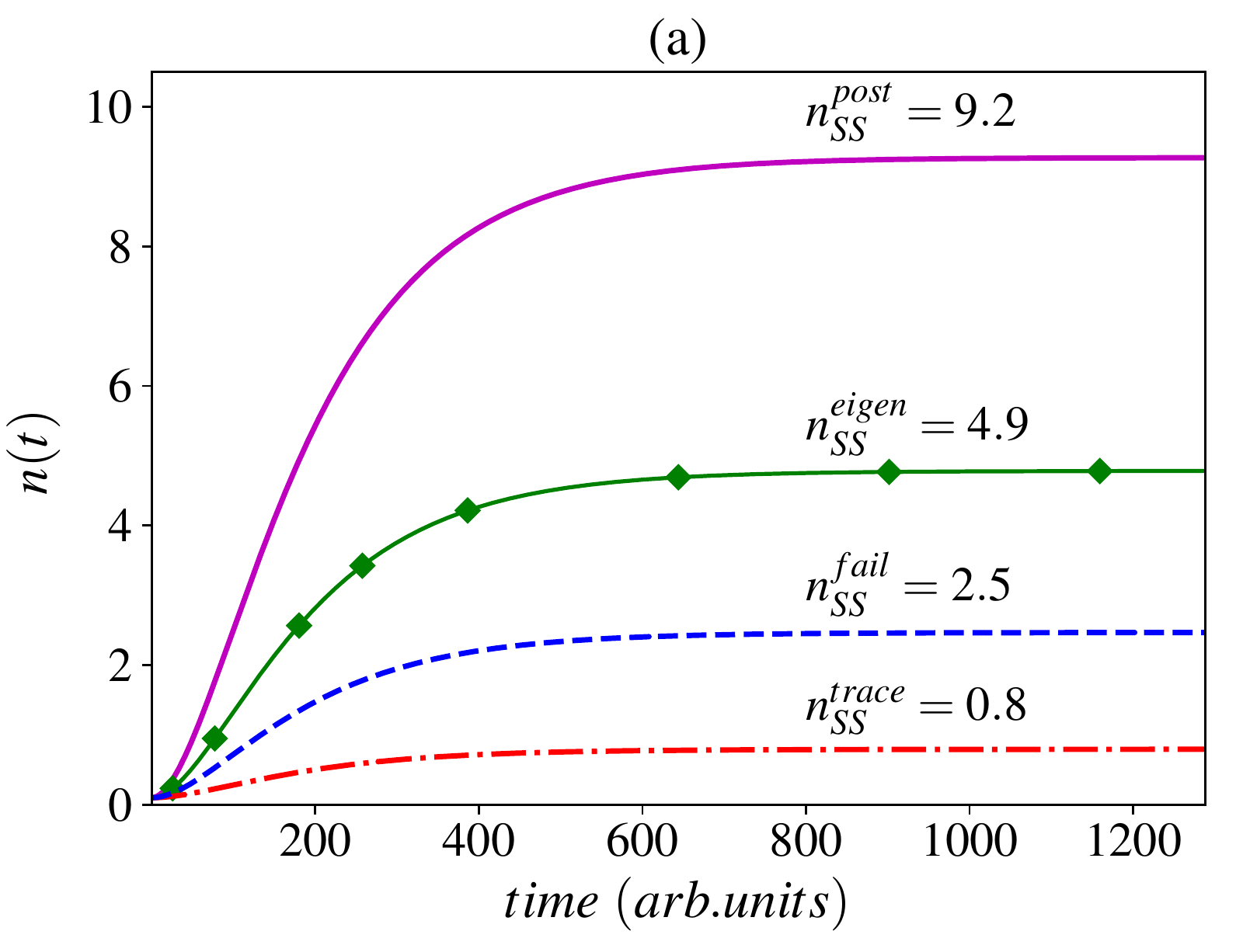}
	\includegraphics[scale = 0.5] {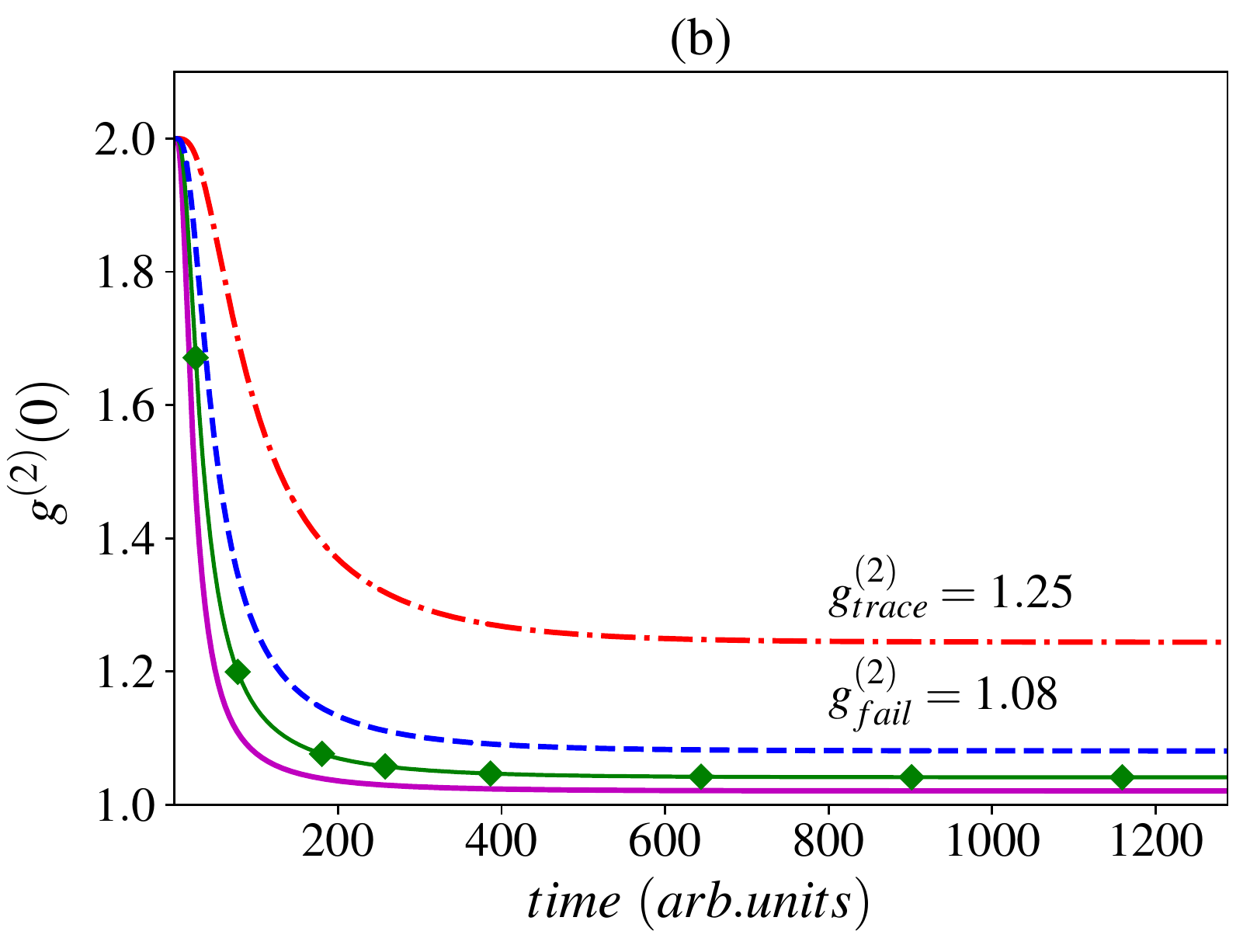}
 	\caption { (Left panel) Time evolution of the average phonon number calculated: i) using heralded post-selection in the state $\ket{\psi_\text{f}}$ results in lasing with the steady-state $\bar{n}_{SS}=9.2$ (magenta line); (ii) using post-selection in the same $\ket{\psi_\text{f}}$ state but with failures, results in lasing with the steady-state $\bar{n}_{SS}=2.5$ (blue dashed); iii) using \textit{spin tracing} which give no lasing (red dot-dashed); iv) using post-selection or \textit{spin tracing} with the initial state $\down$ or $\up$ result in lasing with the steady-state  $\bar{n}_{SS}=4.9$ (green line-diamonds). For the first three cases the initial \textit{unnormalized} state is $\ket{\psi_\text{i}} = 0.4 \up + 0.6 \down$. \\
 		(Right panel) Second-order coherence degree, evidencing the formation of coherent state when $g^{(2)}(0)\to 1$. The result for \textit{spin tracing} operation, i.e. case (iii), is far from a coherent state. The model parameters here are: $\bar{n}_0=0.1$, $\lambda=0.001$, $\kappa=0.014*\lambda$ and $\Delta t=41*\tau$, for $\tau=\pi$.}
 	\label{fig:4}
 \end{figure}

For a more general case, we take into account the unsuccessful post-selective readouts during the full evolution. In order to take full advantage of the Aharonov-Albert-Vaidman \cite{Ah} Weak Value Amplification (WVA) effect, we take the initial and final states nearly orthogonal, in such a way that if $\ket{\psi_\text{i}} = \cos\phi_\text{i}\up + \sin\phi_\text{i} \down$ and $\ket{\psi_\text{f}} = \cos\phi_\text{f}\up - \sin\phi_\text{f}\down$, the successful post selection probability is $P_S\equiv \left |  \langle \psi_\text{f} \ket{\psi_\text{i}} \right |^2$ going to zero.\\
We did the numerical calculations for such a scenario and find the following interesting results: (A) Considering heralded post-selections only (i.e. disregard fully or partially the failures) we can get great amplification, thus a larger stationary phonon number as compared to \textit{spin tracing} or to the case where the final state is $\up$ or $\down$ as in Fig. (\ref{fig:nmed}a), for the same initial conditions. (B) If the failures are included in the maser ME, i.e. using the projector: $P_S \ket{\psi_\text{f}} \bra{\psi_\text{f}} + (1-P_S) \ket{\psi_{\text{f}\bot}} \bra{\psi_{\text{f}\bot}} $, where $\ket{\psi_{\text{f}\bot}}$ is orthogonal to $\ket{\psi_\text{f}}$, then the amplification effect is reduced as compared to (A). Nevertheless, if there is lasing, then the steady-state phonon number, $\bar{n}_{SS}$, is higher than for the case of \textit{spin tracing} operation for the same initial state. Also, it is possible to have a situation of high mechanical losses when the lasing occurs for the post-selections with failures while with \textit{spin tracing} we get no lasing at all. In Fig. (\ref{fig:4}) we illustrate the results of the cases (A) and (B), where we choose the following \textit{unnormalized} initial $\ket{\psi_\text{i}} = 0.4 \up + 0.6 \down$ and final $\ket{\psi_\text{f}} = 0.9 \up - 0.1 \down$ states which give the probability of successful post-selection of $P_S=0.21$, where we see that WVA produces roughly twice the number of phonons as compared to post-selection or \textit{spin tracing} with the initial state $\up$ or $\down$, see cases (i) and (iv) in Fig. (\ref{fig:4}a). \\
In the case when the initial spin state is $\up$ or $\down$, then \textit{spin tracing} or post-selection including failures give the same steady-state phonon number, since the initial state is an eigenstate of the interaction Hamiltonian. However, if the mechanical losses are high (higher than the gain), then the heralded post-selection shows a real advantage, i.e. the WVA effect stimulates the lasing, as shown in Fig.(\ref{fig:4}a). Therefore, in our model we show how the phonon lasing may be stimulated by a protocol as heralded control of the post-selection, similar to \cite{Rao}.
   
    \begin{figure}[t]
    \includegraphics[scale = 0.5] {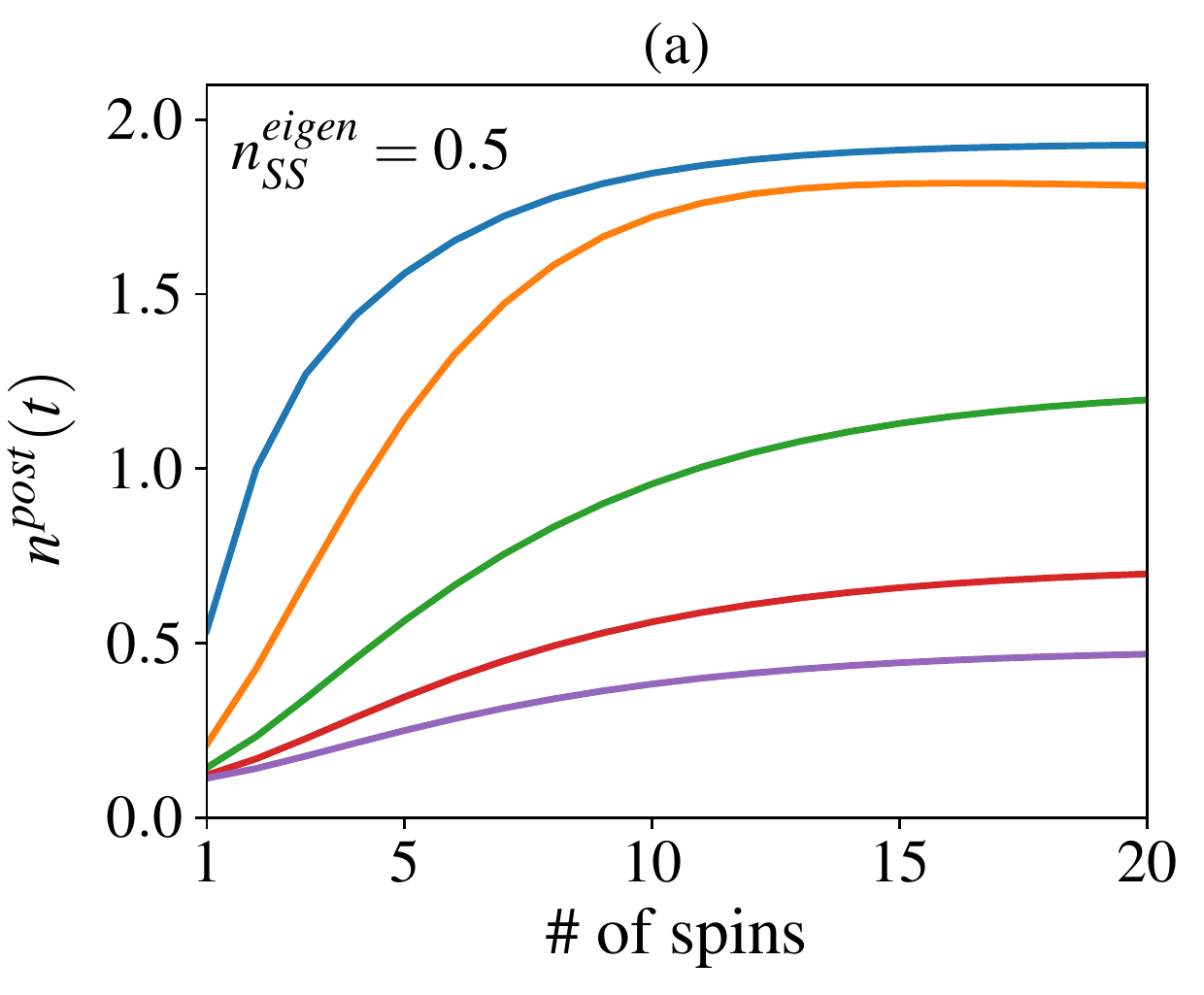}
 	\includegraphics[scale = 0.5] {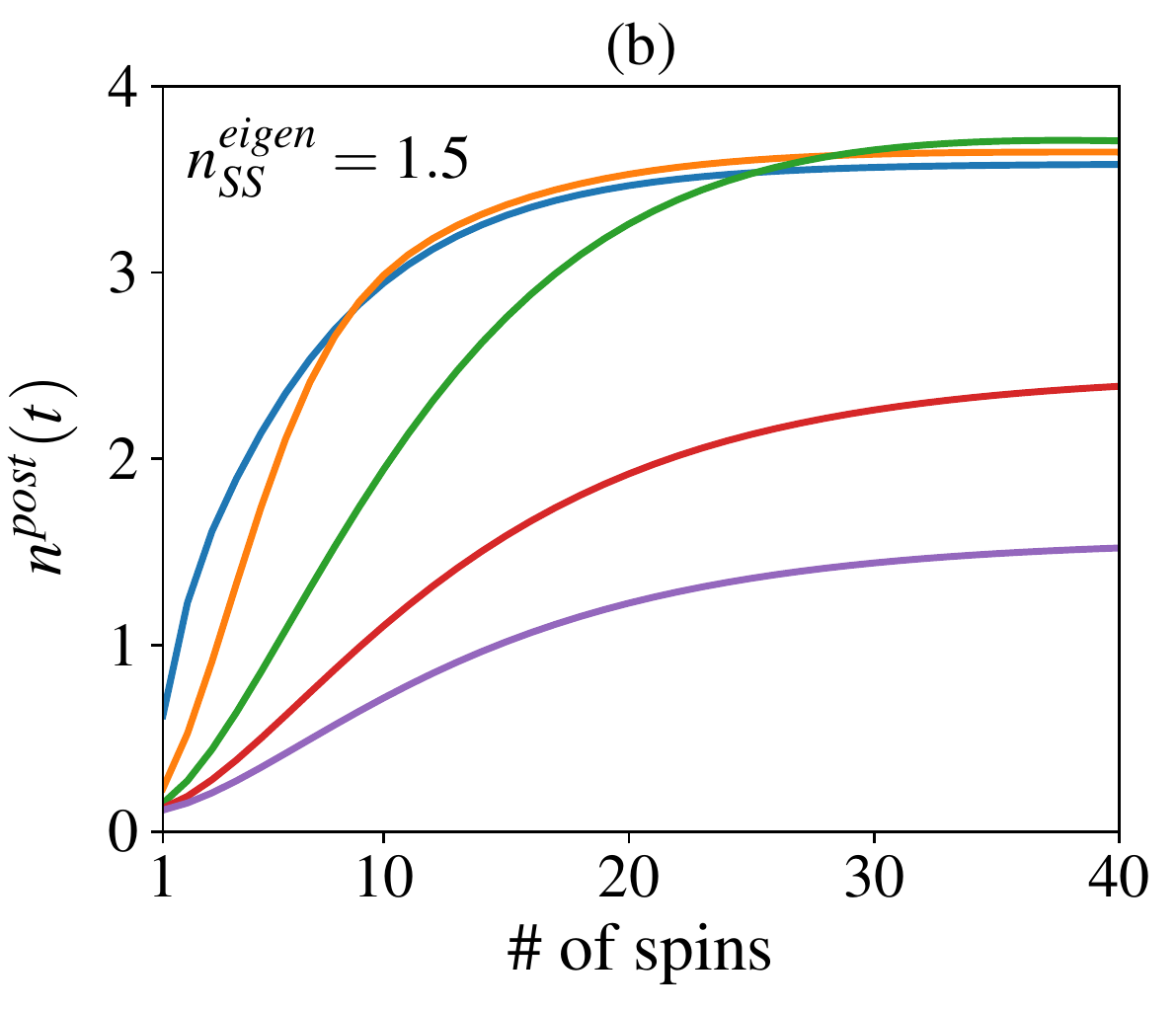}
	\includegraphics[scale = 0.5] {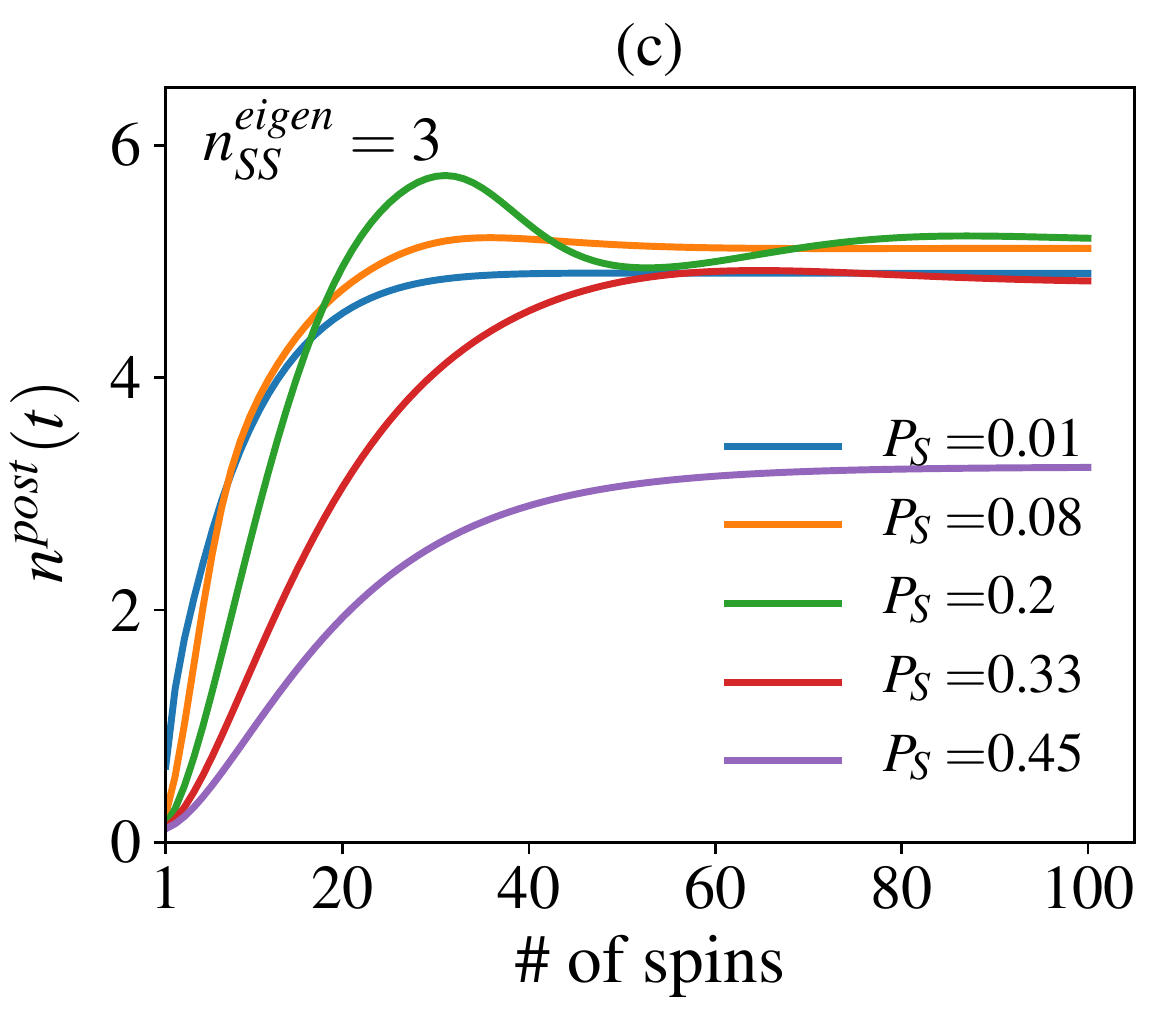}
	\includegraphics[scale = 0.5] {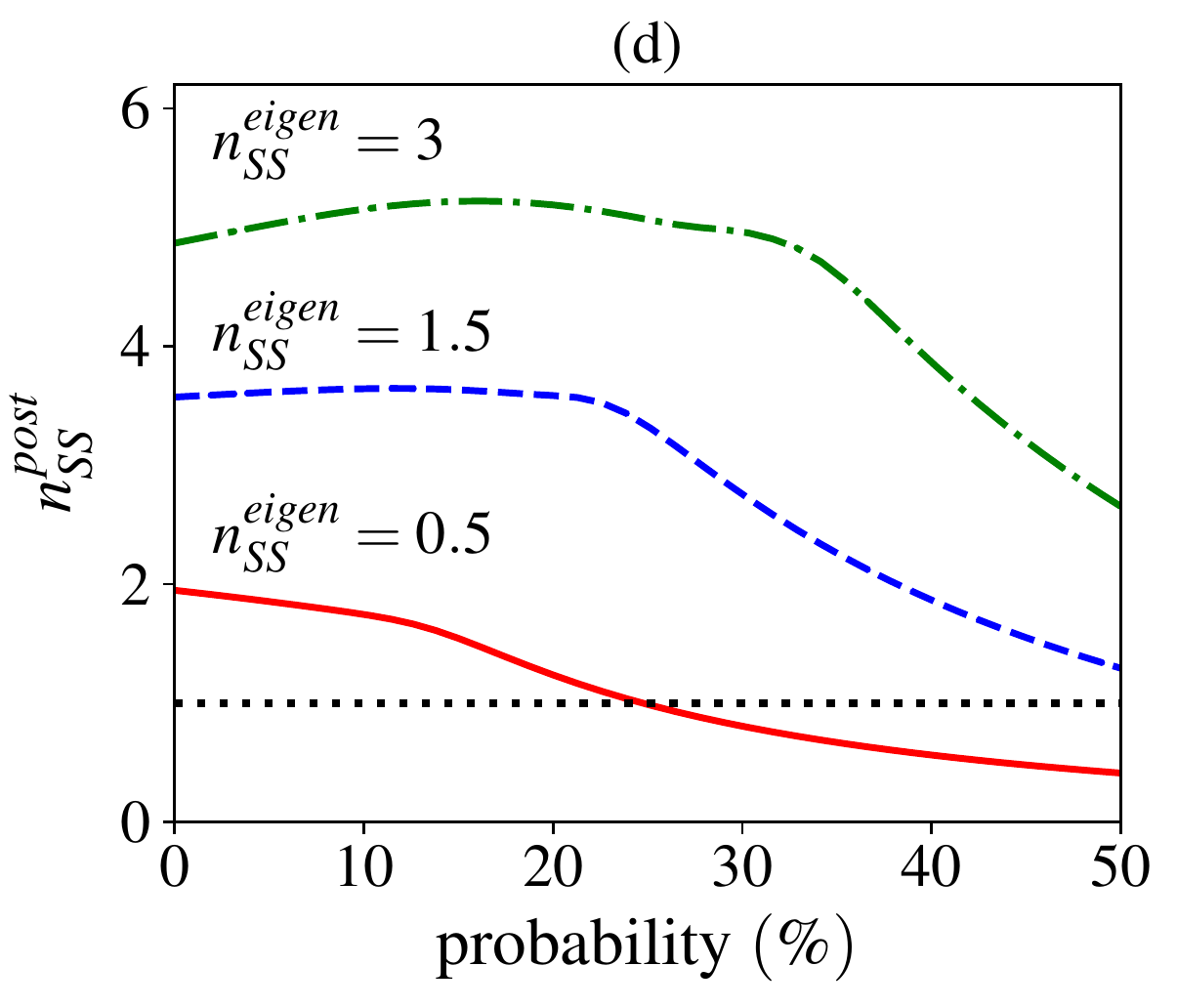}
 	\caption { (a-c) Phonon number as function of the number of spins calculated for different post-selection success probabilities ($P_S$ in legend) and several values of the steady-state phonon number, $n_{SS}^{eigen}$, obtained by \textit{spin tracing} operation for the initial eigenstate of $\hat{\sigma}_z$. (d) The steady-state phonon number vs. post-selection probability. The parameters here are:  $\bar{n}_0=0.1$, $\lambda=0.06$, $\Delta t=35*\tau$, $\tau=\pi$ and the preselected spin state for the post-selection is $\ket{\psi_\text{i}} = (\up + \down)/\sqrt{2}$.}
 	\label{fig:5}
 \end{figure}
 
\section{EXPERIMENTAL FEASIBILITY}
As an experimental case study we suggest the prototype with the physical characteristics similar as in \cite{Rabl, Rao}. For example, a nanomechanical resonator (cantilever), with a magnetic tip (atomic force microscope) at one end, may oscillate at the frequency, $\omega_m \sim 5$ MHz and be coupled via the magnetic field gradient to the electronic spin of the defect in the single nitrogen vacancy center, e.g. see Fig. 1 in \cite {Rabl}. In such setup the spin-mechanics interaction could be considered in the regime from ultra-weak ( $\lambda_0\sim 100$ Hz, \cite {Rao}) to strong ($\lambda_0 \sim 100$ kHz, \cite {Rabl}) coupling, and hence the interaction constant considered by us, i.e. $\lambda_0/ \omega_m \sim 10^{-3}$ belongs to a weak coupling regime. For the modern experimental conditions as low temperatures in the $mK$ regime and high vacuum environment, the quality factor ($Q$) of the resonator can reach a value of $\sim 10^5$, so the dimensionless phonon damping, $\kappa/\omega_m =1/Q \sim 10^{-5}$, as we considered in our numerical calculations, see Fig.(\ref{fig:nmed}). Regarding the spin losses as dephasing and relaxation processes, the corresponding damping rates in the recent experiments are low \cite{Rabl} as compared to the cavity damping, and under these conditions we omitted these losses in our theoretical model.
Finally, in order to witness the phonon lasing, it is possible to implement the readout of the phonon statistics (coherence) as in the experimental proposal \cite{Cohen}, where an analog of Hanbury Brown and Twiss interferometry is used to measure the correlations in the emitted phonons by the single-photon detection.

\section{Discussion}
In summary, we have proposed the model of a phonon maser in a spin-mechanical system with longitudinal interaction accompanied by pre-selection and post-selection measurements. Our model is based on the setup similar to a micromaser with random atomic injection \cite{Ber, SZ, Orszag}, where a sequence of prepared spins are interacting with a mechanical oscillator and after a choice of post-selection of each spin we demonstrate, both analytically as well as numerically, that the oscillator goes to a steady state with a phonon probability distribution close to a Poissonian and almost perfect coherence. As result we find a thresholdless phonon laser, where the non-linear gain in the dynamics depends crucially on the pre-selected and post-selected spin states.

From the experimental point of view, to achieve phonon lasing based on our proposal, the experimentalist should do the best to realize the pre-selective and post-selective measurements with a high success fidelity. Nevertheless, we have checked the impact of the fidelities of pre-selection and post-selection states on the result of the lasing effect. It is interesting to remark that for an initial spin state chosen as an arbitrary superposition, i.e. $\alpha \up + \beta \down$ with randomly generated $ \alpha$ and $ \beta$ one obtains a similar result as for the pre-selected equilibrium superposition, i.e. $(\up + \down)/\sqrt 2$, with only the difference that the dynamics is a bit slower to reach the lasing steady-state. Nevertheless, if the post-selection on the state $\down$ or $\up$ is substituted by a \textit{partial trace} for the same randomly generated initial state, there is no lasing observed as mentioned before. Another effect occurs when the initial spin state is $\up$ or $\down$, then \textit{spin tracing} or post-selection including failures give the same steady-state phonon number, since the initial state is an eigenstate of the interaction Hamiltonian.
Hence, with this result in mind, it seems that it is sufficient to prepare the spins in one of the basis state in order to obtain lasing within the considered spin-mechanical maser setup. Indeed, such a scenario/opportunity may be considered for a real experiment, particularly for low losses as compared to the gain and so obtaining lasing with no need of post-selection. On the other hand, heralded post-selection procedure can advantageously resolve the situation of high losses, e.g. when the lasing with the \textit{spin tracing} operation is impossible for the optimal initial condition. For such circumstances we propose to choose the initial and final states almost orthogonal, e.g. as in Fig.(\ref{fig:4}), then similar to the WVA effect it is possible to stimulate the lasing with the help of heralded post-selection. Experimentally post-selective measurements have been realized successfully for different problems in several laboratories, e.g. \cite{Feiz, Les, Viza}. For example, concerning the high-fidelity post-selective measurement, some recent theoretical and experimental works have proposed to reduce the probe losses due to wrong post-selection, such as power-recycled weak value (PRWV) metrology, e.g. \cite{Wang} or heralded control of mechanical motion \cite{Rao}. These techniques are designed to improve the protocol of WVA in some physical setups by controlling the events of successful post-selections.

As a practical example, particularly to emphasize the advantages and disadvantages of the post-selection strategy versus the \textit{spin tracing} one, we did some numerical comparison as presented in Fig. (\ref{fig:5}). An important technical issue in our model is to maximize the post-selection probabilities and minimize the number of spins necessary to get lasing. This number in fact depends on several parameters, particularly on the coupling constant, steady-state phonon number (as result of gain vs. losses) and on the success probability of each post-selection measurement. Fortunately the relation between the number of spins and the probabilities can be improved by searching optimally the coupling constant as a result of numerical calculations. Therefore, in Fig. (\ref{fig:5} a-c) we show for the post-selection measurement the relationship between the number of spins and the achievable steady-state phonon number for different success probabilities, which is compared to the values obtained by the \textit{spin tracing} operation for the initial eigenstate of the interaction Hamiltonian. We see the lower the probability and steady-state phonon number, the less spins needed to achieve the steady-state. The efficient way to minimize the number of spins is to have low output phonon number, and so in the case (a) one finds that the steady-state lasing for the post-selection with the probability of $8\%$ may be reached with about $10$ spins or maybe less for better optimized parameters. With the same parameters and the eigenstate strategy there is no lasing, resulting in $\bar{n}_{SS}=0.5 <1$ with about $20$ spins. On the other hand, when the lasing with the eigenstate method becomes possible, i.e. for low losses, then the heralded post-selection strategy can be advantageous if a large amplification is required. In order to get a larger phonon number, one has to increase the number of spins in both strategies. However, as found from the simulations the \textit{spin tracing} uses at least double the number of spins, as compared with the post selection. Furthermore, considering the fidelities of the spin preparation, this factor may even be larger. With respect to the overall success probability, of course for many individual post-selective measurements, in fact any overall probability will decrease drastically - this is the price to pay particularly for WVA approach. However as mentioned above, some protocols of the heralded successful post-selections are used to resolve the problem of the successive post-selective measurements, e.g. \cite{Rao}. We expect such a protocol to be promising for the case of reduced number of spins, e.g. about of $10$ spins.

Finally, we point out that all the analytical results presented in our work were compared to their respective quantities obtained from the numerical simulation of the ME (\ref{eq:ME}) by using the Quantum Toolbox in Python (QuTiP) \cite{Joh}. It is important to notice that the both methods of calculation concur very well, sometimes even superposed as in Fig. (\ref{fig:g2}). However, the numerical simulation is based on the approximation of a reduced Hilbert space. In order to solve numerically the master equation (using QuTiP) we have limited the Fock number of excitations to almost three times the steady-state phonon number, particularly cutting the Fock space at $26$, see Fig. (\ref{fig:nmed}b). In fact, the numerical solution could reach the exact solution for the large enough Hilbert space, however such dynamics increases dramatically the computation resources.

\section*{Acknowledgements}

V.E. and M.O. acknowledge the financial support from Fondecyt Regular No. $1180175$.

\section*{Appendix A: General micromaser model}
To apply the micromaser theory to our spin-mechanical system in the case of many spins (main ingredient for micromaser), one considers that (i) the spins interact with the oscillator in sequence, and (ii) the time that one spin interacts with the mechanics, $\tau$, is much shorter than the total time, $t$, of the oscillator's evolution as well compared to the time, $\Delta t$, determining the relaxation under the thermal reservoir action. The density operator of the mechanics after a total time $t$, during which the oscillator interacted with $k$ spins can be written as $\hat{\rho}_m^{(k)}(t) = \hat{M}^k(\tau) \hat{\rho}_m(0)$ \cite{Ber, SZ, Orszag}. Another important ingredient of the laser effect is the the pump mechanism. To model the pump in our proposal let's consider that the spins are approaching to interact with the oscillator at the rate $r$ and that the probability for $k$ spins which are successfully post-selected in the desired state and so contributing to the gain effect, is calculated as $P(k) = C_{Kk} p^{k} (1-p)^{K-k}$, where
$C_{Kk} = K!/k!(K-k)!$, $p$ is the probability for a given spin to be successfully post-selected, and $K$ is the total number of spins involved in the lasing process (i.e., $0 \leq k \leq K$). Therefore, the average number of spins contributing to the gain is $\langle k \rangle = pK$. As a particularity of the micromaser model, the parameter $p$ plays an important role by introducing the effect of the statistics of pumping, with the limit $p\rightarrow0$ (considered in this work) corresponding to random pumping and  $p\rightarrow1$, to  uniform pumping. The latter case does not apply here, since post-selective measurements are probabilistic events with, in fact, very small successful probabilities as in theories and experiments of WVA \cite {Hal, Car, Coto, Ah, Mag}.

Therefore, the density operator of the mechanics, averaged over $k$ successful post-selective measurements, evolves in time as following \cite{Orszag}
\begin{equation}
\hat{\rho}_{m}(t)=\sum_{k=0}^{K} P(k)
\hat{\rho}_{m}^{(k)}(t)=\left\{1 + p[\hat{M}(\tau)-1] \right\}^{K} \hat{\rho}_{m}(0),  \tag{A1}
\label{eq: A1}
\end{equation}
with $K=rt/p$. 

To get the full dynamics for $\hat{\rho}_m(t)$, one computes the derivative of Eq.\ (\ref{eq: A1}) with respect to time and subsequently expand the result in terms of $p (\hat{M}(\tau) - 1)$ up to the second order, so one gets
\begin{equation}
\dot{\hat{\rho}}_{m}(t) = \frac{r}{p}
\ln\left\{1+p[\hat{M}(\tau)-
1]\right\}\hat{\rho}_{m}(t) \simeq r [\hat{M}(\tau)-1]\hat{\rho}_{m}(t)
-\frac{r p}{2}[\hat{M}(\tau)-1]^{2}\hat{\rho}_{m}(t). \tag{A2}
\label{eq: A2}
\end{equation}

\section*{Appendix B: Derivation of phonon distribution}
In terms of Glauber's $P-$representation, the density matrix is $\hat{\rho}_m(t)=\int d^{2}\beta P(\beta,\beta ^*,t) \ket{ \beta} \bra{ \beta}$.
Using the displacement operator, one gets $\hat{M}(\tau)\hat{\rho}_m(t)=\int d^{2}\beta P(\beta,\beta ^*,t)\ket{ -\beta -\eta \lambda} \bra{ -\beta -\eta \lambda} $.

By the standard technique to convert a Master Equation into a Fokker-Planck second order differential equation \cite{SZ} with the loss term $\mathcal{L}\hat{\rho}_m(t) \Longrightarrow \frac{\kappa}{2}(\frac{\partial }{\partial \beta }\beta +\frac{\partial }{\partial \beta ^{\ast }}\beta ^{\ast })P+\kappa \bar{n}_0 \frac{\partial ^{2}P}{\partial \beta \partial \beta ^*}$, we get the time dependent Fokker-Planck equation  
\begin{equation}
 \kappa P+\frac{\partial P}{2\partial \beta }(\kappa\beta -4\lambda r)+\frac{\partial P}{2\partial \beta ^{\ast }}(\kappa \beta ^{\ast }-4\lambda r)+\kappa \bar{n}_0 \frac{\partial ^{2}P}{\partial \beta \partial \beta ^{\ast }}=\frac{\partial P}{\partial t}. \tag{B1}
 \label{eq:B1}
 \end{equation}

In the following, assuming a solution of the type $P=\exp \left[ a(t)+b(t)\beta +c(t)\beta ^{\ast }+d(t)\beta \beta ^{\ast } \right] $ and an initial Gaussian distribution $P(\beta ,\beta ^{\ast },0)=\frac{1}{\pi \varepsilon }\exp (-\frac{\mid \beta -\beta _{0}\mid ^{2}}{\varepsilon })$ one obtains
\begin{eqnarray*}
	b(t)=c(t)=\frac{4\lambda r}{\kappa \bar{n}_0 }(1-\exp [-\kappa t/2])+\frac{\beta_{0}}{\bar{n}_0}\exp [-\kappa t/2], \text{    } 
	d(t)=-\frac{1}{\bar{n}_0 (1-\exp [-\kappa t])+\varepsilon \exp [-\kappa t]}.
\end{eqnarray*}

For an initial thermal distribution $\varepsilon =\bar{n}_0 $, $\beta _{0}=0$, we get $P(\beta,\beta^*,t)=\frac{1}{\pi\bar{n}_0}\exp [-\frac{\mid \beta -\beta_{1}\mid ^{2}}{\bar{n}_0}]$ with $ \beta _{1}=\frac{4\lambda r}{\kappa}(1-\exp [-\kappa t/2]) $.

Therefore the probability of having $n$ phonons is calculated by using the $ P(\beta,\beta^*,t)$ function as following
\begin{equation} 
 \hat{\rho}_{m_{ n,n}}(t) = \int \text{d}^2 \beta P(\beta,\beta^*,t)\vert \inner{n}{\beta} \vert^2=\frac{1}{\pi \bar{n}_0 n!} \int \text{d}^2 \beta \vert\beta\vert^{2n} e^{-\vert\beta\vert^2 - \frac{\vert\beta-\beta_1\vert^2}{\bar{n}_0}},  \tag{B2}
 \label{eq:B2}
\end{equation}

\vspace{10pt}

\textit{Average phonon number.} Using the definition $\left\langle \hat{n}(t) \right\rangle \equiv \text{Tr}\{\hat{n}\hat{\rho}_m(t)\}$ one has 
$\left\langle \hat{n}(t) \right\rangle=\frac{1}{\pi \bar{n}_0}\int d^{2}\beta \mid \beta \mid
^{2}\exp (-\frac{\mid \beta -\beta _{1}\mid ^{2}}{\bar{n}_0})$.

After a straightforward calculation, we get the final expression
\begin{equation}
 \left\langle \hat{n}(t) \right\rangle  = \bar{n}_0+\frac{16 \lambda^2 r^2}{ \kappa^2} \big ( 1- \exp[-\kappa t/2]  \big)^2. \tag{B3}
  \label{eq:B3}
\end{equation}


\end{document}